\documentclass[%
 reprint,
 amsmath,amssymb,
 aps,prl
]{revtex4-2}

\usepackage{graphicx}
\usepackage{dcolumn}
\usepackage{bm}
\usepackage{braket}
\usepackage{mathrsfs}
\usepackage{siunitx}
\usepackage{nicefrac}
\usepackage{soul,xcolor}

\begin{document}

\title{Quantisation and its breakdown in a Hubbard-Thouless pump}

\author{Anne-Sophie Walter}
\email[These authors contributed equally.]{}
\author{Zijie Zhu}
\email[These authors contributed equally.]{}
\author{Marius Gächter}
\author{Joaquín Minguzzi}
\author{Stephan Roschinski}
\author{Kilian Sandholzer}
\author{Konrad Viebahn}
\email[]{viebahnk@phys.ethz.ch}
\author{Tilman Esslinger}
\affiliation{Institute for Quantum Electronics \& Quantum Center, ETH Zurich, 8093 Zurich, Switzerland
}



\maketitle

\textbf{
Geometric properties of waves and wave functions can explain the appearance of integer-valued observables throughout physics~\cite{xiao_berry_2010}.
For example, these `topological' invariants describe the plateaux observed in the quantised Hall effect and the pumped charge in its dynamic analogon, the Thouless pump~\cite{thouless_quantization_1983,kitagawa_topological_2010,oka_floquet_2019}.
However, the presence of interparticle interactions can profoundly affect the topology of a material, invalidating the idealised formulation in terms of Bloch waves.
Despite pioneering experiments in solid state systems~\cite{nuckolls_strongly_2020,yang_quasi-two-dimensional_2022}, photonic waveguides~\cite{jurgensen_quantized_2021}, and optical lattices~\cite{schweizer_spin_2016,tai_microscopy_2017}, the study of topological matter under variation of inter-particle interactions has proven challenging~\cite{rachel_interacting_2018}.
Here, we experimentally realise a topological Thouless pump with fully tuneable Hubbard interactions in an optical lattice and observe regimes with robust pumping, as well as an interaction-induced breakdown.
We confirm the pump's robustness against interactions that are smaller than the protecting gap, which holds true for both repulsive and attractive Hubbard $U$.
Furthermore, we identify that bound pairs of fermions are responsible for quantised transport at strongly attractive $U$, supported by measurements of pair fraction and adiabaticity.
For strong repulsive interactions, on the contrary, topological pumping breaks down.
Yet, we can reinstate quantised pumping by modifying the pump trajectory while starting from the same initial state.
Our experiments pave the way for investigating interacting topological matter~\cite{rachel_interacting_2018} including edge effects~\cite{irsigler_interacting_2019} and interaction-induced topological phases~\cite{ke_multiparticle_2017,lin_interaction-induced_2020,kuno_interaction-induced_2020,bertok_splitting_2022}.}

\begin{figure*}[t!]
\includegraphics[width=1\textwidth]{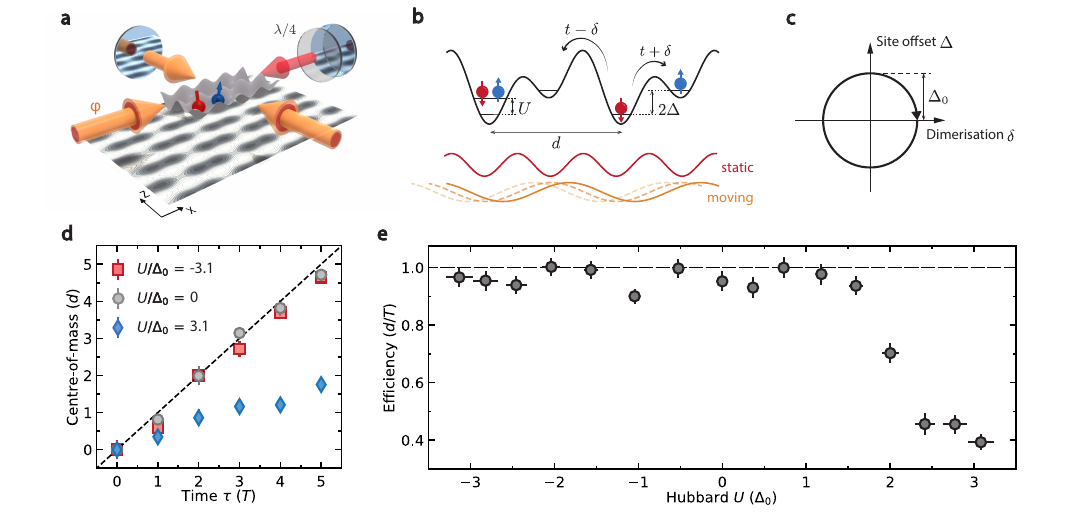}
\caption{\textbf{Topological pumping in the interacting Rice-Mele model} 
\textbf{a}, schematic of the dynamical optical superlattice setup.
The interfering lattice (yellow) is imbalanced along the $x$-direction, leading to a movement of the `long' lattice with respect to the `short' lattice (non-interfering, red arrows) when ramping the phase $\varphi$ of the incoming light.
The running wave component is due to a rotated polarisation of the retro-reflected laser beam ($\lambda/4$-plate). 
The standing wave in $y$-direction is not shown for clarity. $d=\lambda = \SI{1064}{nm}$ is the size of one unit cell. 
\textbf{b}, the resulting lattice structure along $x$ corresponds to the interacting Rice-Mele Hamiltonian (\ref{eqn:RM}).
\textbf{c}, sketched pumping trajectory in the parameter space spanned by site offset $\Delta$ and dimerisation $\delta$.
$\Delta_0$ corresponds to half of the gap in the ionic Hubbard model ($\delta=0)$.
\textbf{d}, measured in-situ centre-of-mass position of the fermionic cloud within five pumping cycles for $U/\Delta_0=\left\{-3.1(2),0,3.1(2)\right\}$ (red, grey, and blue data points, respectively).
Attractive and non-interacting atoms exhibit quantised pumping (black dashed line), while the movement of the repulsive cloud is strongly reduced.
Data points and error bars correspond to mean and standard error of at least eight individual measurements.
\textbf{e}, measured pumping efficiency (fitted slopes of \textbf{b}, averaged over pumping direction) as a function of Hubbard $U$.
Nearly quantised pumping efficiency persists for weakly interacting atoms (both attractive and repulsive) and strongly attractive interactions up to $|U| = 3.1(2)\Delta_0 = 9.2(3)t$.
In the strongly repulsive regime the topological pumping breaks down.
Error bars in $y$-direction correspond to the propagated error estimated from the uncertainty of the fit and those in $x$-direction to the propagated error from lattice fluctuations.
All measurements in this figure were taken at a fixed period of $T = 41.5(1.5) \hbar /t$.}
\label{fig:1}
\end{figure*}

\begin{figure}[b!]
\includegraphics[width=0.5\textwidth]{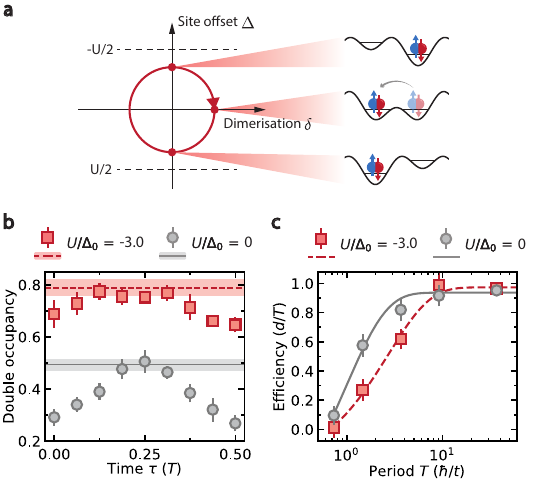}
\caption{\textbf{Quantised pumping of pairs in the attractive Rice-Mele model} 
\textbf{a},
In the strongly attractive regime the pumping mechanism is a result of the tunnelling of pairs of fermions.
\textbf{b}, measured double-occupancy fraction over half a pumping cycle for $U/\Delta_0=0$ (grey points) and $U/\Delta_0=-3.0(1)$ (red squares). 
The large fraction of double occupancies and its small modulation over a pumping cycle for $U/\Delta_0=-3.0(1)$ compared to $U=0$, supports the picture of pair pumping in the strongly attractive regime.
The solid grey and dashed red line indicate the maximum attainable double-occupancy fraction given by our lattice loading scheme.
Each data point and error bar corresponds to the mean and standard error of six individual measurements split equally between pumping directions.
\textbf{c}, adiabatic timescale of the topological pump for non-interacting and strongly attractive atoms. 
The measured efficiency is plotted versus pumping period in units of tunnelling times for $U = -3.0(2)\Delta_0 = -9.2(3)t$ (red squares) and $U=0$ (grey points). 
Data points correspond to the fitted slopes of the c.m.~drift over two pumping cycles averaged over at least 9 iterations and pumping direction.
The data point at $T= 36.5 \hbar/t$ is taken from the data set for Fig.~\ref{fig:1}e at $U/\Delta_0 = -3.1(2)$.
Error bars correspond to the propagated error estimated from the uncertainty of the fit.}
\label{fig:2}
\end{figure}

Ultracold quantum gases provide a versatile platform for investigating topological phenomena~\cite{cooper_topological_2019,zhang_topological_2018,mukherjee_crystallization_2022}, in which atoms take on the role of mobile charges.
Although atoms are electrically neutral, effective magnetic fields can be generated via periodic modulation.
However, the simultaneous presence of interactions and periodic driving often leads to detrimental energy absorption and population of highly excited modes~\cite{oka_floquet_2019,sun_optimal_2020,viebahn_suppressing_2021}.
Most experiments have so far been restricted to the non-interacting regime~\cite{jotzu_experimental_2014,aidelsburger_measuring_2015,nakajima_topological_2016,minguzzi_topological_2022} 
or interactions remained fixed~\cite{lohse_thouless_2016,schweizer_spin_2016}.
Conversely, realising a many-body system with topology and variable interactions is still a challenge, despite significant and ongoing theoretical interest~\cite{niu_quantised_1984,berg_quantized_2011,qian_quantum_2011,grusdt_realization_2014,zeng_fractional_2016,tangpanitanon_topological_2016,li_finite-size_2017,lindner_universal_2017,ke_multiparticle_2017,hayward_topological_2018, nakagawa_breakdown_2018,stenzel_quantum_2019,haug_topological_2019,unanyan_finite-temperature_2020,greschner_topological_2020,lin_interaction-induced_2020,kuno_interaction-induced_2020,chen_simulating_2020,fu_nonlinear_2022,mostaan_quantized_2022,esin_universal_2022,bertok_splitting_2022}.

In our experiment, we create a dynamically tuneable superlattice by overlaying phase-controlled standing waves with an additional running wave component and study topological charge pumping in the periodically driven, interacting Rice-Mele model~\cite{rice_elementary_1982},
\begin{eqnarray}\label{eqn:RM}
    \hat{H}(\tau) &=& - \sum_{j,\sigma}\left[t + (-1)^j\delta(\tau)\right]\left(\hat{c}_{j\sigma}^\dagger \hat{c}_{j+1\sigma} + \text{h.c.}\right) \\
    \nonumber &&+\,\Delta(\tau)\sum_{j,\sigma} (-1)^j \hat{c}_{j\sigma}^\dagger \hat{c}_{j\sigma}+U\sum_{j} \hat{n}_{j\uparrow} \hat{n}_{j\downarrow}~.
\end{eqnarray}
The interactions enter as the Hubbard $U$ for two fermions of opposite spin $\sigma \in \{\uparrow,\downarrow \}$ occupying the same lattice site~$j$.
The fermionic annihilation and number operators are denoted by $\hat{c}_{j\sigma}$ and $\hat{n}_{j\sigma}$, respectively.
Both the bond dimerisation $\delta(\tau)$ and sublattice site offset $\Delta(\tau)$ are sinusoidally varied in time $\tau$ with period $T$, but out of phase with respect to each other.
This cyclic and adiabatic modulation describes a quantum pump, which manifests itself in a drift of the many-body polarisation \cite{vanderbilt_berry_2018}.
For an insulator or a homogeneously filled band of free fermions ($U=0$) this drift is quantised, realising a Thouless pump~\cite{thouless_quantization_1983}, protected by the single-particle gap of the bipartite lattice structure (Eq.~\ref{eqn:RM}, Fig.~\ref{fig:lattice_setup}, and Methods).
During pumping our system remains in the low-energy sector described by Eq.~\ref{eqn:RM}.
For finite interactions ($U\neq 0)$ the Rice-Mele model encompasses a rich many-body phase diagram ~\cite{torio_phase_2001}, including the ionic Hubbard model with maximum site offset $\Delta_0$ and no dimerisation \cite{pertot_relaxation_2014,messer_exploring_2015} as well as the interacting SSH model with maximum dimerisation $\delta_0$ and zero site offset \cite{de_leseleuc_observation_2019}.

The experiments are performed using a balanced spin-mixture $(\uparrow,\downarrow)$ of ultracold potassium-40 atoms in a three-dimensional optical lattice (Fig.~\ref{fig:1}, Fig.~\ref{fig:lattice_setup}, and Methods). 
The total lattice potential comprises interfering laser beams in the $x$--$z$ plane and additional non-interfering standing waves in all three spatial directions, $x$, $y$, and $z$~\cite{tarruell_creating_2012}.
These potentials combine to form one-dimensional superlattices along $x$. 
The phase between the interfering (`long') lattice with respect to the non-interfering (`short') lattice along $x$ is dynamically controlled, inspired by the self-oscillating mechanism of ref.~\cite{dreon_self-oscillating_2022}.
This traces an elliptical path of the Rice-Mele parameters $\delta$ and $\Delta$ around the origin.
In contrast to previous realisations in optical lattices~\cite{lohse_thouless_2016,nakajima_topological_2016}, our setup uses a single laser source at $\lambda = \SI{1064}{nm}$ for all lattice beams, avoiding wavelength-dependent phase shifts in the optical path.
Prior to pumping, we use a loading scheme with an intermediate lattice to increase the fraction of atoms in doubly occupied unit cells.
This fraction is characterised with an independent measurement, achieving approximately $80\%$ in the strongly attractive regime, and around $50\%$ in both the weakly interacting and strongly repulsive regimes (Fig.~\ref{fig:loading} and Methods). 

In a first experiment we track the many-body polarisation by measuring the centre-of-mass (c.m.) position of the atomic cloud within five pumping cycles for varying interaction strengths $U$ (Fig.$~$\ref{fig:1}d).
We choose the lattice such that the single-particle gap is approximately constant over such a pumping cycle and is given by $2\Delta_0$.
Fitting a line to this data yields the efficiency of the pump, which is plotted versus $U$ in Fig.$~$\ref{fig:1}e.
This measurement characterises the topological behaviour of an interacting Thouless pump, allowing us to distinguish three cases.
Firstly, quantised pumping (indicated by the black dashed line) persists for weak interactions ($U\lesssim\Delta_0 = 2.9t$) both in the attractive and in the repulsive situations.
We attribute deviations from unity to fluctuations in the position of our in-situ atomic cloud and residual changes in momentum distribution owing to drifts in filling fraction.
Secondly, this plateau of nearly quantised topological transport, averaging to an efficiency of 0.96(3), extends to large attractive values of $U$ which exceed the single particle gap $2\Delta_0$. 
Increasingly large attractive Hubbard $|U|$ leads to the formation of double occupancies~\cite{micnas_superconductivity_1990}.
Topological pumping then relies on the transport of pairs, in contrast to the standard description of pumping with single atoms~\cite{greschner_topological_2020}.
Importantly, we observe a clear asymmetry between strongly attractive and strongly repulsive interactions.
Beyond $U\simeq +2\Delta_0 = 5.8t$, topological pumping breaks down and its efficiency decreases down to 0.39(3).
For comparison, we perform numerical simulations of the many-body ground state at half-filling with a density matrix renormalisation group (DMRG) algorithm on four to 64 lattice sites and find that the pumping efficiency drops to zero for large $U$ (Fig.~\ref{fig:effi-vs-L-U}).
The remaining pumping efficiency of $0.4-0.5$ for large repulsive interactions in the experimental data is a result of the non-zero fraction of atoms in singly occupied unit cells. This agrees with an independent measurement of the initial state, where $55(7)\%$ of atoms are found in doubly occupied unit cells.
Our observations therefore support the picture that topological pumping breaks down for doubly occupied unit cells at large repulsive $U$.

The observation of pair pumping for strong attractive interactions is substantiated by additional observables, including the evolution of double occupancy fraction (DO) and the timescale for adiabaticity.
We detect the fraction of pairs over half a trajectory for the non-interacting ($U=0$, grey data points) and strongly attractive system ($U/\Delta_0=-3.0(1)$ red data points), as shown in Fig.$~$\ref{fig:2}b.
The solid grey and dashed red line indicate the maximum accessible DO given by our lattice loading, which is equal to the fraction of the initially doubly occupied unit cells determined via an additional measurement (Methods).
In the absence of interactions, the delocalisation of atoms within a unit cell lead to a finite DO at $\tau =0$ and $\Delta = 0$. 
A large negative $U$ gives rise to an increased initial DO.
While the DO increases by more than 0.2 over the course of a cycle for $U=0$ when reaching the maximum site offset $\Delta_0$, the high fraction in the attractive system only increases by half as much.
Thus, we can conclude that the pairs for $U/\Delta_0 = -3.0(1)$ largely remain bound over the pumping cycle.
The residual modulation in DO over half the pumping cycle is also reflected by the DMRG simulations in Fig.~\ref{fig:DO}.
By analogy, quantised pumping should also be possible with repulsively bound pairs~\cite{ke_multiparticle_2017} for $U>0$, which we plan to investigate in the future.

The pumping of pairs also manifests itself in a change of adiabaticity timescale, compared to single atoms.
Generally, the timescale for adiabatic following is determined by the minimum energy gap to the first excited state over a pumping cycle, which in the non-interacting Rice-Mele model corresponds to the second Bloch band of the bipartite lattice.
In the experiment, the transport efficiency for the attractive pairs drops at longer periods compared to $U =0$.
Exponential fits to the data points yield $1/e$ times of $2.7(4)\hbar/t$ for $U/\Delta_0 = -3.0(2)$ (dashed red line) and $1.0(2)\hbar/t$ for $U=0$ (solid grey line).
The increase in adiabaticity timescale indicates that the energy gap becomes smaller in the attractive regime and agrees with the estimate $2t^2/|U| \simeq 0.33(1) t$ (at $\tau = 0$, Fig.~\ref{fig:lattice_setup}) for effective tunnelling of hardcore bosons.

\begin{figure}[t!]
\includegraphics[width=0.5\textwidth]{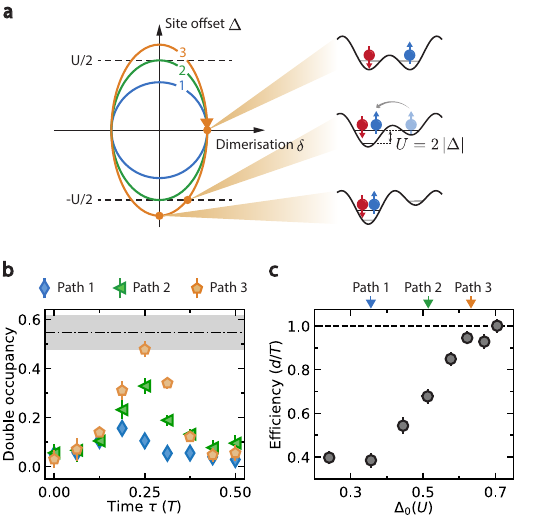}
\caption{\textbf{Quantised pumping via resonant tunnelling for strongly repulsive interactions ($U \simeq 8 t$).} 
\textbf{a}, three pumping trajectories with varying maximum site offset $\Delta_0$ and fixed dimerisation $\delta_0$.
All three paths start in the same state. 
The first trajectory (blue path) prohibits formation of double occupancies and therefore precludes transport.
Trajectories 2 and 3, with larger $\Delta_0$ (blue and green paths), allow for tunnelling between sites of a unit cell when crossing the $\left|\Delta (\tau) \right| = U/2$ line parallel to the dimerisation axis (dashed black lines). 
\textbf{b}, measured double occupancy fraction over half a pumping cycle for the three trajectories shown in \textbf{a}. 
\textbf{c}, pumping efficiencies versus maximum site offset $\Delta_0$ for varying pumping paths. 
The efficiency increases from 0.40(3) to unity as $\Delta_0$ becomes large enough to allow for resonant tunnelling between a double occupancy and localized single occupancies on each site. 
Data points and errors bars are obtained analogously to Fig.$~$\ref{fig:2}b and c.}
\label{fig:3}
\end{figure}

Next we investigate how to recover quantised transport in the strongly repulsive regime where pumping breaks down (Fig.~\ref{fig:1}d).
To that end, we modify the pump trajectory and increase the maximum site offset $\Delta_0$ (path 2 and 3 in Fig.$~$\ref{fig:3}a), compared to the initial trajectory (path 1), while keeping the starting point and the interactions fixed.
Path 1 corresponds to the data point with the same absolute $U$ in Fig.~\ref{fig:1}e ($U = 2.8(1) \Delta_0 = 8.0(3) t$).
As a result of increasing $\Delta_0$, single occupancies and double occupancy become resonantly coupled by tunnelling~\cite{ke_multiparticle_2017}.
Thus, an asymmetric charge distribution within a unit cell becomes energetically allowed. 
This asymmetry manifests in the change in polarisation and is necessary for transport.
We demonstrate this process in our experiment by measuring the double occupancy fraction for all three trajectories shown in Fig.$~$\ref{fig:3}a over half a pumping cycle (for path 1 to 3 $ \Delta_0 = \left\{ 0.35(1), 0.50(1), 0.61(1) \right\} U$ with fixed $U$).
The initial fraction is below 0.1 for all paths considered here, reflecting the identical initialisation.
While the DO for path 1 remains below 0.17, it reaches values of 0.33(3) and 0.48(2) for paths 2 and 3, as the line of $\Delta = U/2$ is crossed and pair formation restored.
For path 3, the measured fraction even reaches the maximum possible value within error (black dashed-dotted line), determined by the initially doubly occupied unit cells (Methods).
The observation is qualitatively consistent with numerical calculations (Fig.~\ref{fig:DO_res}).
The influence of resonant pair formation on transport becomes clear with a measurement of efficiency versus maximum site offset $\Delta_0$ in Fig.$~$\ref{fig:3}c. 
For low values of $\Delta_0$, the pump efficiency is roughly constant at around 0.4. 
Increasing $\Delta_0$ leads to a growth in efficiency up to unity as the resonance condition for tunnelling is fulfilled.

In conclusion we have experimentally characterised the topological properties of interacting Thouless pumps covering the full range of Hubbard $U$, from strongly attractive, through intermediate to strongly repulsive.
Remarkably, we observe a clear asymmetry between large attractive and large repulsive interactions. 
While the robustness of quantised pumping of the former can be explained by an effective hardcore boson picture, the latter experiences a marked breakdown of transport. 
The experimental tools presented in this work also provide a pathway to studying how interactions affect the role of spatial~\cite{rachel_interacting_2018} and temporal disorder, as well as edge physics~\cite{irsigler_interacting_2019}.
Furthermore, our approach could enable topological transport which has no counterpart in the limit $U\rightarrow 0$, leading to novel interaction-induced topological states~\cite{ke_multiparticle_2017, kuno_interaction-induced_2020, lin_interaction-induced_2020,bertok_splitting_2022}.

\textit{Note:} during the review of this manuscript we became aware of a related works~\cite{jurgensen_quantized_2023,leonard_realization_2022}.

\subsection{Acknowledgements}

We would like to thank Jason Ho and Gian-Michele Graf for inspiring and insightful discussions as well as Daniel Malz for comments on the manuscript. We thank Alexander Frank for his contributions to the electronic setup of the experiment. K.V. is supported by the ETH Fellowship Programme. This work was partly funded by the SNF (project no. 182650), NCCR-QSIT, and ERC advanced grant TransQ (project no. 742579).

\subsection{Author contributions}

The data was measured and analysed by A.-S.W., Z.Z., M.G., J.M., and K.V. A.-S.W. and Z.Z. performed the numerical calculations. K.V. and T.E. supervised the work. All authors contributed to planning the experiment, discussions and the preparation of the manuscript.

\subsection{Author information}
The authors declare no competing financial interests. Correspondence and requests for materials should be addressed to K.V. (viebahnk@phys.ethz.ch) and T.E. (esslinger@ethz.ch).


%

\clearpage

\section{METHODS}

\subsection{Experimental sequence}
We start by evaporatively pre-cooling a cloud of fermionic potassium $^{40}$K in the magnetic state $F = 9/2$, $m_F = -9/2$ and confining it to a crossed dipole trap.
We then create a spin mixture of $m_F = \{-9/2,-7/2\}$ and further evaporatively cool it, yielding $47'000(4'000)$ atoms at a temperature of $0.11(3)T/T_F$.
Values in brackets correspond to the standard deviation over all measured data points; the atom number is calibrated within a systematic error of $10\%$.
Atoms are subsequently loaded into a three-dimensional lattice within \SI{200}{ms}.
Using the magnetic Feshbach resonance at 202.1 G, we tune the s-wave scattering length between the atom in the $-7/2$ and $-9/2$ sublevels to be very strongly attractive $a\rightarrow -\infty$.
Subsequently loading the atoms into a shallow chequerboard within \SI{200}{ms} and then deep chequerboard within \SI{10}{ms} leads to a high double occupancy fraction~\cite{tarruell_creating_2012}.
To reach large repulsive interactions ($U/\Delta_0\gtrsim 1$ in Fig.~\ref{fig:1}d, e and Fig.~\ref{fig:3}) we then apply a radio-frequency sweep, transferring the atoms in the $-7/2$ to the $-5/2$ magnetic sublevel and keeping the $-9/2,-7/2$ mixture otherwise.
To reach the final lattice, we ramp the magnetic field to the scattering length yielding the targeted $U$ and split the sites of the chequerboard lattice into two sites along the $x$-direction (Fig.~\ref{fig:loading}).
This loading procedure results in many copies of the ground state of half-filled double wells (one of each spin in one unit cell).
The improved loading procedure results in a larger fraction of atoms in doubly occupied unit cells and larger number of holes, which varies with the interaction strength during the split.

Compared to the previous loading scheme~[arXiv:2204.06561v1], where the atoms were directly loaded into lowest band of the final lattice, this fraction contributes to a larger breakdown signal for repulsive interactions (see the comparison of pump efficiencies versus $U$ in Fig.~\ref{fig:effi-vs-U-compare} and the DMRG calculations versus $U$ in Fig.~\ref{fig:effi-vs-L-U} for different system sizes). Also, this preparation precludes the presence of atoms in the higher band and therefore a more persistant unity pumping efficiency on the attractive side.
The resulting trapping frequencies are $86.1(1.1)$, $77.3(0.8)$, $121.8(1.2)$\SI{}{Hz} in the $x$, $y$, and $z$-directions.

\subsection{Detection methods}
After pumping the system for varying times, 
we either measure the in-situ centre-of-mass~(c.m.) position of our cloud or detect the double occupancy fraction.
\paragraph{Centre of mass position} We detect the in-situ c.m. position of our atomic cloud by taking an absorption image directly after the ramp of the phase, in the presence of the dipole trap, optical lattice and homogeneous magnetic field. 
The conversion from pixel size to lattices sites is done by independently measuring the displacement of the cloud in a lattice with very large gap to the next excited band over 50 cycles in steps of 10.
\paragraph{Double-occupancy fraction}
For the double-occupancy fraction we first freeze the dynamics of the atoms by quenching into a deep cubic lattice within \SI{100}{\mu s}. 
We then sweep the magnetic field over the $-7/2$, $-9/2$ Feshbach resonance and spectroscopically resolve the interaction shift with radio-frequency radiation, transferring atoms in the $-7/2$($-5/2$) state in doubly occupied sites to the $-5/2$($-7/2$) state. 
The Zeeman sublevels are then separated by applying a magnetic field gradient and \SI{8}{ms} time-of-flight~\cite{viebahn_suppressing_2021}.
\paragraph{Fraction of atoms in half-filled unit cells}
For the determination of fraction of atoms in doubly occupied unit cells, we take the sum of double occupancy fraction, singlet and triplet fraction. 
As for the double occupancy measurement, the detection of the latter two start with a freeze ramp into a deep cubic lattice.
Double occupancies are then eliminated by applying two (one) consecutive Landau-Zener sweeps, transferring the atoms in the $-7/2$ ($-5/2$) state to the $-3/2$ state: doubly occupied sites then host a very-short lived $-3/2$, $-9/2$ mixture which is lost from the trap.
A magnetic gradient leads to an oscillation between the two populations, and the extrema yield the singlet and triplet fraction.
To measure these, the lattice is ramped to a checkerboard configuration which merges adjacent sites. 
As a consequence of the Pauli exclusion principle, triplets are then converted to one atom in the lowest band and one in the higher, whereas singlets form double occupancies in the lowest band.
These single or double occupancies are detected with the same method as previously described for the double occupancies.
For normalization of the fraction of atoms in half-filled unit cells we take the number of atoms $N$ from the same measurement as the one done to assess the number of double occupancies.

\subsection{Optical lattice}
The lattice is made up of four retro-reflected beams at a wavelength of $\SI{1064}{nm}$. The non-interfering beams in $x$, $y$, and $z$-direction create a cubic lattice to which the interfering beams in the $x$--$z$ plane superimpose a chequerboard lattice.
The resulting potential as seen by the atoms is given by
\begin{equation}\label{eqn:potential}
\begin{split}
     V(x,y,z) = &- V_{\text{X}} I_{\text{self}} \cos^2 (kx + \vartheta / 2) \\
     &- V_{\text{Xint}} I_{\text{self}} \cos^2(kx)\\
     &- V_{\text{Y}} \cos^2(ky) \\
     &- V_{\text{Z}} \cos^2(kz) \\
    &- \sqrt{V_{\text{Xint}} V_{\text{Z}}} \cos(kz) \cos(kx + \varphi) \\
    &- I_{\text{XZ}} \sqrt{V_{\text{Xint}} V_{\text{Z}}} \cos(kz) \cos(kx - \varphi)~,
\end{split}
\end{equation}
where $k=2 \pi / \lambda$. The lattice depths $[V_{\text{X}},V_{\text{Xint}},V_{\text{Y}},V_{\text{Z}}]$ used in this paper are given by $[6.02(4) , 0.37(3) , 14.98(3) , 17.0(3) ] E_R$, measured in units of recoil energy $E_R = h^2/2m\lambda ^2$, where $m$ the mass of the atoms.
The phase $\varphi$, which is the relative phase between the incoming lattice beams in $x$- and $z$-direction, governs the depth and the relative position of the chequerboard with respect to the square lattice.
The angle $\vartheta$, defining the relative position between the one-dimensional sinusoidal lattice formed by $V_{\text{X}}$ and that formed by $V_{\text{Xint}}$, is controlled by the difference in light frequency of the two beams.
We calibrate $\vartheta$ to $1.000(2)\pi$ by minimising the double occupancy during splitting of a chequerboard into a dimerised lattice at $U=0$.
The imbalance factors $I_{\text{self}}$ and $I_{\text{XZ}}$ are due to the $\lambda / 4$ waveplate in the retro-path (Fig.~\ref{fig:1} and Fig.~\ref{fig:lattice_setup}).
The factor $I_{\text{XZ}}$ plays a crucial role in our pumping scheme which is based on sliding a varying chequerboard lattice over a square lattice.
The sliding is achieved by ramping the relative phase $\varphi$ which is stabilized using a locking scheme, detailed in the next section.
Without the imbalance (i.e.~$I_{\text{XZ}} = 1$), as was the case in our previous work~\cite{viebahn_suppressing_2021}, the phase $\varphi$ would enter as an overall amplitude $\cos(\varphi)$.
However, in case of $I_{\text{XZ}} < 1$ the interference terms proportional to $\sqrt{V_{\text{Xint}} V_{\text{Z}}}$ in Eq.~\ref{eqn:potential}  acquire a $\varphi$-dependent position, explaining the ability to slide the chequerboard using $\varphi$. 
We rotated the $\lambda / 4$ waveplate such that the incoming, linearly polarised light is rotated by $26^{\circ}$ after passing the plate twice.
This results in imbalance factors of $I_{\text{self}} = 0.98(2)$ and $I_{\text{XZ}} = 0.81(2)$, which are independently calibrated using lattice modulation spectroscopy.

The Rice-Mele parameters in Eq.~\ref{eqn:RM} are  calculated via the basis of maximally localised Wannier states, spanning the space of solutions to the single particle Hamiltonian with potential Eq.~\ref{eqn:potential}.
Overlap integrals between these Wannier states yield the relevant tight-binding tunnelling elements, on-site energies, as well as interactions $U$.
The values of $\Delta$, $\delta$, and $t$ are plotted in Fig.~\ref{fig:lattice_setup}c as function of $\varphi \in [0, 2\pi]$.
Typical parameters are $\Delta_0 \simeq 3.0t$, $\delta_0 \simeq 1.5t$, leading to small variations of the single-particle band gap between $1.8\,\Delta_0$ and $2.0\,\Delta_0$ over one period.
Sinusoidal fits to this data simplify the theoretical description; the resulting fit parameters are listed in Table~\ref{tbl:rm_params}.
Due to the strong confinement along $y$ and $z$ the tunnellings along those directions $t_\text{Y,Z}$ are below \SI{20}{Hz} over the whole pump cycle.
The onsite interaction $U$ is given by \SI{995}{Hz} for a reference scattering length of 100 Bohr radii, which varies by about $3 \%$ over the pump cycle, and the interaction between neighbouring sites is always below \SI{50}{Hz}.

\subsection{Phase lock}

Topological pumping is realised by shifting the interference phase $\varphi$ in time.
The scheme for controlling the phase $\varphi$ is illustrated in Fig.~\ref{fig:phaselock}, taking the $x$-direction as an example.
The setup is replicated on the $z$-axis, which is not shown in Fig.~\ref{fig:phaselock} for clarity.
Active stabilisation of the light phase is necessary since the optical fibre introduces significant phase noise.
In short, the back-reflection from the optical lattice forms a Michelson-interferometer together with a reference beam, which does not pass through an optical fibre.
In this manner, the absolute phase of the lattice can be measured, assuming a perfectly stable reference arm.
We shift the phase of the lattice beam by using the frequency modulation input (`FM in') of a Rohde \& Schwarz (`RS') function generator (SMC100A) creating the RF-frequency for the acousto-optic modulator (AOM).
A small frequency shift will result in a phase shift of the laser beam at the position of the atoms (red cloud in Fig.~\ref{fig:phaselock}).
We additionally correct for small deviations to the absolute phase by shifting the phase of the output of the 'RS' generator to the AOM.
The set-point of the phase can now be varied in two different ways: For long pumping periods (longer than \SI{5}{ms}) an Arbitrary-Waveform-Generator (AWG) (Keysight 33500B) generates a sawtooth signal as the set-point of the phase lock, which results in a linear phase ramp.
For short pumping cycles (less than \SI{10}{ms}) the bandwidth of the phase lock is not large enough to follow the set-point.
In this case the AWG creates a square signal, which is added to the feedback signal from the phase lock before the frequency modulation input using a power splitter.
The square waveform after integration also results in a linear phase shift of the lattice beam.
For example, a frequency shift of \SI{400}{Hz} on the RF signal of the AOM leads to a pumping slope of $\Delta \varphi/\Delta\tau = 2\pi/\SI{5}{ms}$.

\subsection{DMRG calculations}

Numerical results of pumping efficiency and double occupancy dynamics presented are calculated with density matrix renormalisation group (DMRG) using the TeNPy python package~\cite{hauschild_efficient_2018}.
The polarization and double-occupancy dynamics shown in Fig.~\ref{fig:effi-vs-L-U}, Fig.~\ref{fig:DO} and Fig.~\ref{fig:DO_res} are calculated using open boundary conditions (OBC), where we assume $L=64$ and half filling (one of each spin in one unit cell). Throughout the calculation, we have selected a maximum bond dimension of $\chi=100$. The tight-binding parameters used in the simulation are identical to those used in the corresponding experiments. The polarisation, i.e., the centre of mass of the ground state $\ket{\Psi(t)}$ is defined by 
$$P_{\rm{open}}(t)=\frac{1}{L} \sum_{\sigma} \sum_{j=0}^{L-1} \bra{\Psi(t)}(j-j_{0})\hat{n}_{j\sigma}\ket{\Psi(t)},$$
and the double occupancy fraction $\mathcal{D}$ is defined as the fraction of atoms on doubly occupied lattice sites
$$\mathcal{D} = \frac{2}{N} \sum_{j}\braket{\hat{n}_{j\uparrow}\hat{n}_{j\downarrow}}.$$
where N is the total atom number.

\subsection{Data availability}
All data files are available from the corresponding author on request. Source Data for Figs 1, 2, 3 and Extended Data Figs 4, 5, 6 are provided with the online version of the paper.

\clearpage

\setcounter{figure}{0} 
\setcounter{equation}{0} 

\renewcommand\thefigure{ED\arabic{figure}} 

\begin{figure}[htbp]
    \centering
    \includegraphics[width=0.45\textwidth]{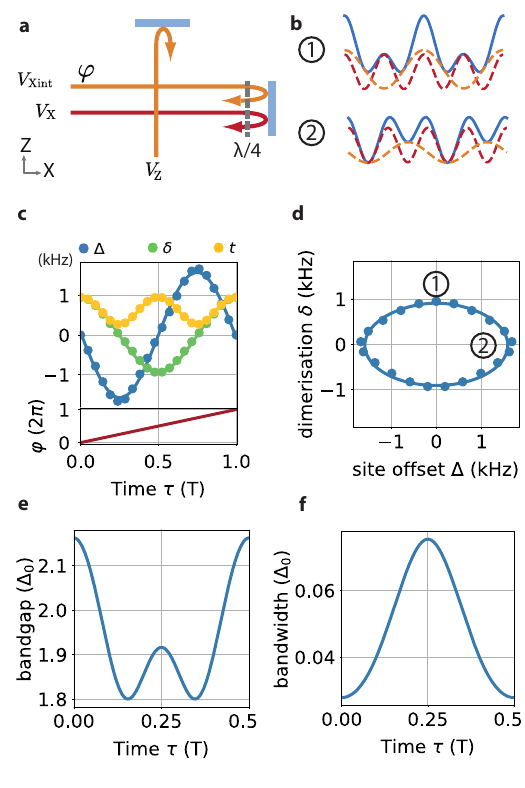}
    \caption{\textbf{Optical lattice setup and Rice-Mele parameters.} \textbf{a}, schematic of the optical setup in the $x$--$z$ plane. The $\lambda / 4$ waveplate produces an intensity imbalance between the different contributing beams in such a way that the phase $\varphi$ can be used to move the chequerboard over the square lattice, realising pumping. \textbf{b}, idealised cut through the lattice potential in $x$-direction, corresponding to the two points along the pump cycle shown in \textbf{d}. The site-offset case (2) is exaggerated for clarity.
    \textbf{c}, tight-binding parameters during one pump cycle. The phase $\varphi$ (defined in Eq.~\ref{eqn:potential}) is ramped from 0 to $2\pi$. In the main text the time dependence of the average tunnelling $t$ has been dropped for clarity.
    In contrast to previous realisations of the Rice-Mele pump with cold atoms~\cite{lohse_thouless_2016,nakajima_topological_2016}, the site offset $\Delta$ and dimerisation $\delta$ follow a sinusoidal waveform over a pump cycle (solid lines corresponds to the fitted sinusoid and the corresponding parameters are summarized in table~\ref{tbl:rm_params}.
    The single-particle gap is dominated by the dimerised tunneling at $\tau = 0$ and by the site offset at $\tau = T/4$.
    \textbf{d}, elliptical trajectory of $\delta$ and $\Delta$ over one pump cycle. The solid line corresponds to the fitted curves in \textbf{c}.
    Single-particle band gap (\textbf{e}) and bandwidth (\textbf{f}) in units of $\Delta_0$ over a pump half-cycle (the second half is symmetric).}
    \label{fig:lattice_setup}
\end{figure}

\clearpage

\begin{figure}
\includegraphics[width=0.5\textwidth]{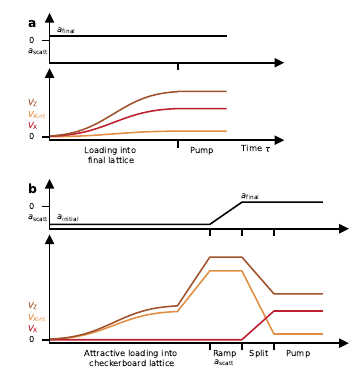}
\caption{\textbf{Lattice and interaction ramps for two different loading schemes.}
(a) The `conventional' lattice ramps with final depths $[V_{\text{X}},V_{\text{Xint}},V_{\text{Y}},V_{\text{Z}}]$ of $ [5.40(5) , 0.09(2) , 15.02(6) , 17.04(8) ] E_R$ are made up of s-shaped ramps at the final scattering length for the targeted $U$.
(b) The improved scheme includes an intermediate ramp into a deep checkerboard lattice at strongly attractive scattering lengths to maximise the fraction of atoms in half-filled unit cells.
Intended interactions are then reached by ramping the magnetic field strength, for the $-9/2,-7/2$ mixture, or a ramp and a prior RF pulse for the $-9/2,-5/2$ mixture.
Splitting the single cells of the checkerboard pattern in two then yields the final lattice depths $[6.02(4) , 0.37(3) , 14.98(3) , 17.0(3) ] E_R$.}
\label{fig:loading}
\end{figure}

\clearpage

\begin{figure}
\includegraphics[width=0.5\textwidth]{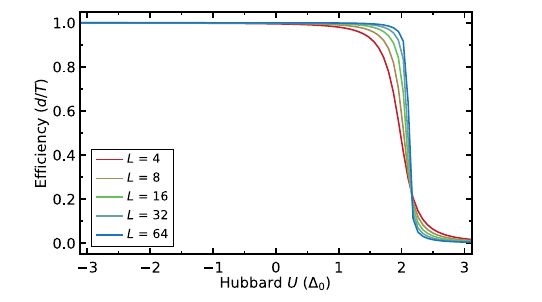}
\caption{\textbf{Effect of finite system size.} Pumping efficiency vs. Hubbard $U$ at different system size $L$ in unit of lattice site, calculated with DMRG, assuming half-filling in OBC. The tight-binding parameters are chosen to be the same as in Fig.~\ref{fig:1}e. Smaller system sizes, which we expect from our loading scheme,  only slightly change the position and steepness of the transition between the quantised and break down regime versus $U$. }
\label{fig:effi-vs-L-U}
\end{figure}

\clearpage

\begin{figure}
\includegraphics[width=0.5\textwidth]{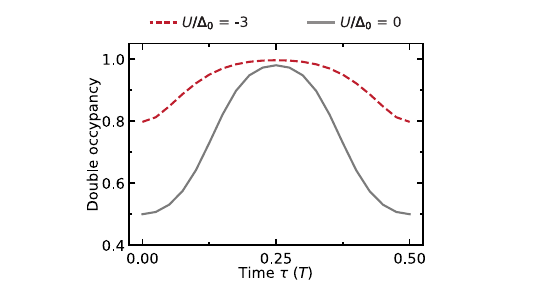}
\caption{\textbf{Numerical simulation of double occupancy fraction over half a pumping cycle with different Hubbard interactions $U$.}
The numerical simulations assume $L = 64$ lattice sites in OBC and half-filling. The tight-binding parameters are chosen to be the same as in Fig.~\ref{fig:2}b. Compared to the simulations, we record an overall lower double occupancy in the experiment (Fig.~\ref{fig:2}b) as a result of an overall average filling lower than one half.}
\label{fig:DO}
\end{figure}

\clearpage

\begin{figure}
\includegraphics[width=0.5\textwidth]{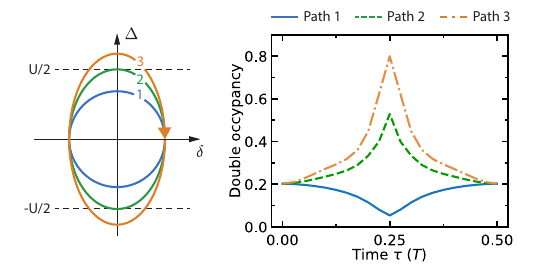}
\caption{\textbf{Numerical simulation of double occupancy (DO) fraction over half pumping cycle with different pumping trajectories.}
The numerical simulations assume $L = 64$ lattice sites in OBC and half-filling. The maximum site offset $\Delta_{0}$ for path 1,2,3 and the Hubbard $U$ are chosen to be the same as in Fig.~\ref{fig:3}b.
The experimental data (Fig.~\ref{fig:3}b) shows an overall lower double occupancy as a result of an average filling lower than one half. For path 1 the ground state simulation and experimental data do not exhibit the same behaviour versus $\tau$. We attribute the difference to: (i) imperfect double occupancy detection around $\tau=0$ and $\tau=T/2$. Due to a large tunnelling rate, the ramp to a square lattice for the DO detection (Methods) is not fast enough to completely freeze the dynamics of the atoms, which yields a slightly lower DO fraction than its actual value, (ii) at $\tau=T/4$, the system can be characterized by an ionic Hubbard model, which exhibits gapless spin excitations. Therefore, the ideal adiabatic following of the instantaneous ground state is hindered, resulting in a lower than expected DO fraction after $\tau=T/4$.}
\label{fig:DO_res}
\end{figure}

\clearpage

\begin{figure}
\includegraphics[width=0.5\textwidth]{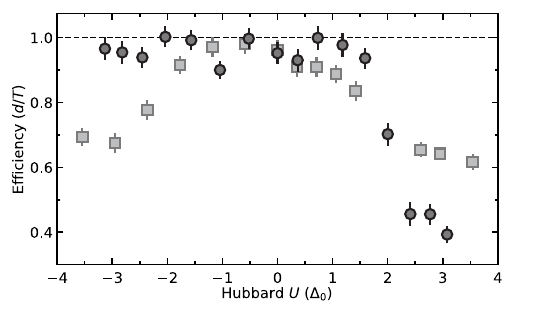}
\caption{\textbf{Pumping efficiency vs $U$ for two different loading schemes and lattices.} The pump efficiency with the improved loading scheme (dark circles, $[\Delta_{0},\delta_{0},t] = [1750, 900, 625]\,\SI{}{Hz}$) exhibits a more pronounced plateau for attractive interactions, compared to conventional lattice loading (grey squares, $[\Delta_{0},\delta_{0},t] = [847, 460, 403]\,\SI{}{Hz}$). The significant improvement of the signal on the attractive side is a result of both the new loading scheme, as well as a different final lattice configuration. In the improved lattice, the energy gap is roughly twice as large (\SI{3.6}{kHz} compared to \SI{1.84}{kHz} in the conventional lattice) which reduces the higher band population during the loading. Likewise, the breakdown of transport on the repulsive side shows a steeper decline beyond a critical $U$, compared to conventional lattice loading. Error bars correspond to the propagated error estimated
from the uncertainty of the linear fit.
}
\label{fig:effi-vs-U-compare}
\end{figure}

\clearpage

\begin{figure}[htbp]
    \centering
    \includegraphics[width=0.5\textwidth]{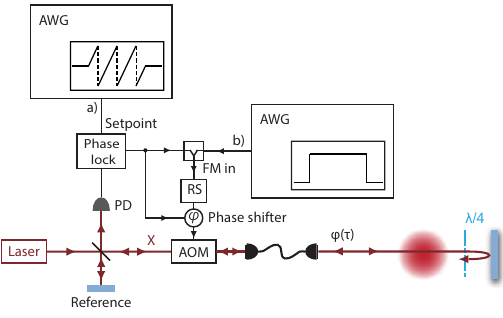}
    \caption{\textbf{Schematic of the phase control.} The back-reflected lattice beam forms a Michelson interferometer together with the reference path before the optical fibre. A linear increase in the interference phase can be realized either by linearly ramping the set-point (a) or by using a square waveform as input for the frequency modulation, which will also result in a linear phase ramp (b). }
    \label{fig:phaselock}
\end{figure}

\clearpage

\setlength{\tabcolsep}{0.5em}
\begin{table}
\def\arraystretch{1.5}
\begin{tabular}{ ccccc }
\hline
\hline
parameter & offset & ampl. & freq. & phase offset\\
 & B [Hz] & A [Hz] & $\nu$ & $\kappa$ \\
\hline
$t$ & 625 & 340 & 2 & $\pi / 2$ \\ 
$\Delta$ & 0 & 1750 & 1  & $\pi$\\
$\delta$ & 0 & 900 & 1 & $\pi / 2$ \\
\hline
\end{tabular}
\caption{Rice-Mele parameters for the fitted sines in Fig.~\ref{fig:lattice_setup}. The parameters correspond to the expression \mbox{$B + A \sin(2\pi \nu \tau / T + \kappa)$}, where $\tau$ is time and $T$ is the pump period.} 
\label{tbl:rm_params}
\end{table}

\end{document}